\documentclass[12pt]{iopart}

\usepackage{amssymb}
\usepackage{graphicx}
\usepackage[usenames,dvipsnames]{color} % for color
\usepackage[usenames,dvipsnames]{xcolor} % for color
\usepackage[normalem]{ulem} % for strikethrough
\newcommand{\beq}{\begin{equation}}
\newcommand{\eeq}{\end{equation}}
\newcommand{\bea}{\begin{eqnarray}}
\newcommand{\eea}{\end{eqnarray}}
\newcommand{\eps}{\varepsilon}

\newcommand{\mkrm}[1]{}           % use this to hide what has been removed

\begin{document}

\title{First application of Fayans functional to deformed nuclei}

\author{S V Tolokonnikov$^{1,2}$, I N Borzov$^{1,3}$, M Kortelainen$^{4,5}$, Yu S Lutostansky$^1$,
and E E Saperstein$^{1,6}$}
\address{$^1$Kurchatov Institute, 123182 Moscow, Russia}
\address{$^2$ Moscow Institute of Physics and Technology,  Dolgoprudny, Russia}
\address{$^3$ Joint Institute for Nuclear Research, 141980 Dubna, Russia}
\address{$^3$ Department of Physics, P.O. Box 35 (YFL), University of Jyv\"askyl\"a, FI-40014 Jyv\"askyl\"a, Finland}
\address{$^4$ Helsinki Institute of Physics, P.O. Box 64, FI-00014 University of Helsinki, Finland}
\address{$^6$ National Research Nuclear University MEPhI, 115409 Moscow, Russia}

\hskip 1.7 cm E-mail: tolkn@mail.ru, ibor48@mail.ru, markus.kortelinen@jyu.fi,\\
.\hskip 2.5 cm lutostansky@yandex.ru, saper43\_7@mail.ru

\begin{abstract}
 First calculations for deformed nuclei with the Fayans functional
are carried out for the uranium and lead  isotopic chains. The ground state deformations and
deformation energies are compared  to Skyrme-Hartree-Fock-Bogolyubov   HFB-17 and HFB-27 functional
results. For the uranium isotopic chain,  the  Fayans functional property predictions are rather
similar to HFB-17 and HFB-27 predictions. However, there is a disagreement for the lead isotopic
chain. Both of the Skyrme  HFB functionals lead to predictions of rather strong deformations for the
light Pb isotopes, which does not agree with the experimental data on charge radii and magnetic
moments of the odd Pb isotopes. On the other hand, the Fayans functional leads to the prediction  of a
spherical ground state for all of the lead isotopes, in accordance with the data and the results known
from the literature obtained with the Gogny D1S force and the SLy6 functional as well. The deformation
energy curves are calculated and compared against those derived from four Skyrme functionals---SLy4,
Sly6, SkM* and UNEDF1---for the $^{238}$U nucleus and several lead-deficient Pb isotopes. In the first
case, the Fayans functional result is rather close to SkM* and UNEDF1 ones, which---in particularly
the latter---describe the first and second barriers  in $^{238}$U rather well. For the light lead
isotopes, the Fayans deformation energy curves are qualitatively close to those derived from the SLy6
functional.
\end{abstract}

\maketitle

\section{Introduction}
 A long-standing goal of the low-energy nuclear theory community is to
have a unified theoretical framework, applicable to nuclear structure and reactions. Presently, due to
computational limitations, {\it ab initio} approaches are applicable to light  or medium mass
closed-shell nuclei only. Therefore, microscopical theories which use effective forces with
phenomenological parameters are usually applied to describe the entire nuclear chart. Nuclear density
functional theory (DFT)  provides  the most popular such models.  In the framework of nuclear DFT,
complex many-body correlations are encoded  into  the energy density functional (EDF), constructed
from  the nuclear densities and currents. Historically, since the work by Vautherin and Brink
\cite{HF-VB}, the Hartree-Fock (HF) method  with the effective Skyrme forces has become very popular
in nuclear physics. From the very beginning, the  Skyrme HF method was aimed at calculating global
properties of nuclei, such as the binding energy and neutron and proton density distributions. A
little later the HF method with the effective Gogny force was suggested \cite{Gogny} and successfully
applied to the same objects as the Skyrme HF method had been applied too. In addition to these
approaches, relativistic mean-field (RMF) model methods have  also been employed in nuclear physics;
see  \cite{RMF} and references therein.  In fact, it was quite soon realized that these methods had a
rather strong correspondence with DFT methods employed e.g. in atomic physics.  Indeed, during the
last few decades, mean-field methods in the framework of HF  and Hartree-Fock-Bogolyubov (HFB) theory
have been widely used in nuclear physics \cite{(RingSchuck),[Ben03]}. The HFB, a method suitable for
superfluid nuclei with pairing correlations, is a generalization of the HF approach, which allows
particles and holes to be treated on  an equal footing.

The use of density-dependent effective interactions is  a  common feature of these  mean-field
approaches. When Skyrme and Gogny effective forces are written as a form of EDF, a rather simple
ansatz for density dependence is assumed. Schematically it reads \beq E_0^{\rm int}[\rho]=\int {\cal
E}(\rho({\bf r})) d^3r=\int \frac {a \rho^2} 2 \left(1+\alpha \rho^{\sigma}\right) d^3r,\label{E0SHF}
\eeq where $ \rho({\bf r})$ is the matter density and $a,\;\alpha$, and $\sigma\leq 1$ are parameters.
For brevity, we omit for a time the isotopic indices and do not discuss the spin-orbit and other
``small'' terms of the effective force. As will be discussed later, Fayans functional  has more
sophisticated density dependence \cite{Fay1,Fay3,Fay4,Fay5,Fay}. Recently, the density dependence of
Skyrme-like EDFs has been enriched by utilizing density matrix expansion techniques
\cite{[Dob10],[Geb10],[Kai10],[Car10],[Geb11],[Sto10]}.

The parameters of Skyrme forces, as well as Gogny and RMF models, have been typically adjusted to the
experimental data on nuclear binding energies and charge radii. Many optimization schemes also use
data on single-particle levels and fission properties,  together with  other observables  and
pseudo-observables. Because data on  nuclei that are very neutron rich are scarce, and were especially
scarce at the time when some of the older Skyrme parameterizations were adjusted, some of the
isovector parameters may have larger uncertainties. The best description of nuclear masses (the
root-mean-square deviation from the respective experimental values being smaller than 600 keV) was
attained with the HFB-17  EDF by the Brussels-Montreal Collaboration  \cite{HFB-17,site}. This result
was achieved, however, by including some phenomenological corrections on top of the mean field.

The Fayans functional \cite{Fay1,Fay3,Fay4,Fay5,Fay}   used in this work assumes a rather
sophisticated  density dependence which can be schematically written as \beq {\cal E}(\rho) = \frac{a
\rho^2} 2 \frac{1+ \alpha \rho^{\sigma}}{1+ \gamma \rho^{\sigma}}, \label{Fay1} \eeq  where $\gamma$
is one of the EDF parameters.  The use of the bare mass, i.e. $m^*=m$, is another peculiarity of the
Fayans functional. Both of these features of the Fayans approach are connected to the self-consistent
theory of finite Fermi systems (TFFS) \cite{KhS}.

Up to now, all applications of the Fayans functional were limited to spherical nuclei. In addition to
the aforementioned investigations, they included the analysis of charge radii \cite{Sap-Tol}, of the
magnetic \cite{mu1,mu2} and quadrupole \cite{QEPJ,QEPJ-Web} moments in odd nuclei,  of characteristics
of the first $2^+$ excitations in even semi-magic nuclei \cite{BE2,BE2-Web}  and of beta-decay
\cite{beta} as well. Recently, single-particle spectra of magic nuclei have been analyzed
\cite{Levels}. In all of the aforementioned cases, a reasonable description of the data was
achieved---better as a rule than that achieved in analogous Skyrme HFB calculations.

 It is worth discussing briefly another new recently developed approach, initially known as the
Barcelona-Catania-Paris EDF and latterly as the Barcelona-Catania-Paris-Madrid (BCPM) EDF. The volume
part of the BCPM EDF is found from the infinite nuclear matter Brueckner--Hartree--Fock approach, by
using a realistic free $NN$ potential. The infinite nuclear matter equation of the state  (EOS)  is
then approximated within a good accuracy by two polynomials: one for the isoscalar  and one for the
isovector components. In addition to this, the EDF contains a finite range term and a term for the
spin-orbit force. This approach was initially formulated in \cite{BCPM0}, whereas the final form of
the corresponding EDF was developed in \cite{BCPM1}. Later, the properties of the BCPM EDF have been
developed and investigated further \cite{BCPM2,BCPM3,BCPM4,BCPM5,BCPM6,BCPM7,BCPM8}. In~\cite{BCPM6},
masses of 579 nuclei were fitted with the rms deviation of 1.58 MeV. The charge radii were also
described with a very good accuracy.  To some extent, the BCPM EDF is similar to the Fayans EDF.
Indeed, the volume part of the FaNDF$^0$ functional, used in the present work, was adjusted
\cite{Fay5} to the Friedman and Pandharipande nuclear and neutron matter EOS \cite{Pandh}. Similarly,
the volume part of the BCPM, given by Baldo {\it et. al.} \cite{BCPM0,BCPM1}, was also adjusted to the
calculated nuclear matter EOS. The use of a bare mass, that is $m^*=m$, is another common feature
 of these two EDFs.

The aim of this work is to apply, for the first time, the Fayans functional in the study of deformed
nuclei. The principle goal is to study deformation properties of Fayans functional for a selected set
of isotopic chains. This will pave the way for more comprehensive studies with the Fayans functional
across the nuclear chart.  In the present work, the general finite range structure of the Fayans EDF
\cite{Fay1,Fay3,Fay4} is localized to a form which is closer to the Skyrme EDFs. This allows us to
employ the computer code HFBTHO~\cite{code}, developed for Skyrme EDFs, with some modifications. This
article is organized as follows. In section 2, the  path  from the self-consistent TFFS to the Fayans
functional is outlined. In section 3, the version FaNDF$^0$ \cite{Fay5} of the Fayans functional is
briefly described.  Section 4 presents the  results for U and Pb isotopes calculated with the set
\cite{Fay5} of FaNDF$^0$ parameters. Section 5 contains our conclusions.

\section { Self-consistent TFFS and  the  Fayans functional}
The self-consistent  TFFS \cite{KhS} is based on the general principles of TFFS \cite{AB1}
supplemented with the condition of self-consistency in the TFFS among the energy-dependent mass
operator $\Sigma({\bf r_1},{\bf r_2};\eps)$, the single-particle Green's function $G({\bf r_1},{\bf
r_2};\eps)$, and the effective NN interaction ${\cal U}({\bf r_1},{\bf r_2},{\bf r_3},{\bf
r_4};\eps,\eps')$ \cite{Fay-Khod}.

  This approach starts from the quasiparticle mass operator $\Sigma_q({\bf r_1},{\bf r_2};\eps)$
which, by definition \cite{AB1}, coincides with the exact mass operator $\Sigma$ at the Fermi surface.
In the mixed coordinate-momentum representation the operator $\Sigma_q({\bf r},k^2;\eps)$ depends
linearly on the momentum squared   $k^2$ and the energy $\eps$ \cite{AB1,KhS},\beq \Sigma_q({\bf
r},k^2;\eps)= \Sigma_0({\bf r}) + \frac 1 {2m \eps^0_{\rm F}} \, {\bf k} \, \Sigma_1({\bf r}){\bf k} +
\Sigma_2({\bf r}) \frac {\eps} {\eps^0_{\rm F}}, \label{sigmaq}\eeq where $\eps^0_{\rm F}=(k^0_{\rm
F})^{\Large 2}/2m$ is the Fermi energy of nuclear matter, $k^0_{\rm F}$  being the corresponding Fermi
momentum.   By definition, we have     \beq \Sigma_0({\bf r})=  \left. {\Sigma({\bf
r},k^2;\eps)}\right|_0, \label{Sig0} \eeq  \beq \Sigma_1({\bf r})= \eps^0_{\rm F}\left.\frac {\partial
\Sigma({\bf r},k^2;\eps)} {\partial \eps_k}\right|_0, \label{Sig1} \eeq \beq \Sigma_2({\bf r})=
\eps^0_{\rm F}\left.\frac {\partial \Sigma({\bf r},k^2;\eps)} {\partial \eps}\right|_0, \label{Sig2}
\eeq where $\eps_k=k^2/2m$ and the subscribe  `0'  means that the energy and momentum variables are
taken at the Fermi surface.  Thus, the component $\Sigma_2$ determines the $Z$-factor: \beq Z({\bf
r})=\left ( 1- {\Sigma_2({\bf r})} / {\eps^0_{\rm F}} \right)^{-1}, \label{Zfac1}\eeq whereas the
inverse effective mass is \beq \frac m {m^*({\bf r})} = \frac {\left ( 1+ \Sigma_1({\bf r}) /
{\eps^0_{\rm F}} \right)}{\left ( 1- \Sigma_2({\bf r}) / {\eps^0_{\rm F}} \right)}. \label{efmass}\eeq
Usually, the quantity inverse to the numerator is called the `$k$-mass', and the denominator, the
 `$E$-mass'.

  The wave functions $\psi_{\lambda}({\bf r})$ which diagonalize the quasiparticle Green
function $G_q=(\eps-\eps_k-\Sigma_q)^{-1}$ obey the following equation: \beq \left( \Sigma_0({\bf r})
-\frac 1 {2m \eps^0_{\rm F}} \nabla \Sigma_1({\bf r})\nabla+ \Sigma_2({\bf r}) \frac {\eps_{\lambda}}
{\eps^0_{\rm F}} \right)\psi_{\lambda} = \eps_{\lambda} \psi_{\lambda}. \label{psiq}\eeq They are
orthonormalized with the weight, \beq \int d {\bf r}\,\psi^*_{\lambda}({\bf r})\,\psi_{\lambda'}({\bf
r}) \,\left( 1- \Sigma_2({\bf r})/\eps^0_{\rm F} \right) = \delta_{\lambda \lambda'} \label{norm}.\eeq

 The Lagrange formalism was used in  Ref.~\cite{KhS},  with  the  quasiparticle Lagrangian $L_q$ being
constructed in such a way that the corresponding Lagrange equations coincide with equation
(\ref{psiq}).

In  the doubly magic nuclei, which are non-superfluid, the Lagrangian density ${\cal L}_q$, with
$L_q=\int d{\bf r}{\cal L}_q({\bf r})$, depends on three sorts of densities $\nu_i({\bf
r}),\;i=0,1,2$: \beq \nu_0({\bf r})= \sum n_{\lambda} \psi^*_{\lambda}({\bf r})\psi_{\lambda}({\bf
r}), \label{nu0}\eeq \beq \nu_1({\bf r})= -\frac 1 {2m \eps^0_{\rm F}}\sum n_{\lambda}
\nabla\psi^*_{\lambda}({\bf r})\nabla\psi_{\lambda}({\bf r}), \label{nu1}\eeq
 \beq \nu_2({\bf r})= \frac 1 {\eps^0_{\rm F}}\sum
n_{\lambda} \eps_{\lambda} \psi^*_{\lambda}({\bf r})\psi_{\lambda}({\bf r}), \label{nu2}\eeq where
$\eps_{\lambda}$ and $n_{\lambda}$ are the quasiparticle energies and occupation numbers, and
 $n_{\lambda}=(0,1)$. Evidently, one gets \beq   \nu_0({\bf r})= Z({\bf r}) \rho ({\bf r}), \label{rho} \eeq
 where the density $\rho ({\bf r})$ is normalized to the total particle number.
The relation between $\nu_1({\bf r})$ and the Skyrme density $\tau ({\bf r})$ is more complicated
\cite{KhS}. The density $\nu_2({\bf r})$ has no analogue in the Skyrme HF theory.

  The components $\Sigma_i$ of the mass operator (\ref{sigmaq}) can be found from the
interaction Lagrangian  $L'_q[\nu_i]$ as follows: \beq \Sigma_i= \frac {\delta L'_q} {\delta \nu_i}.
\label{sigmai} \eeq   The simplest ansatz for the quasiparticle Lagrangian which involves the momentum
and energy dependence effects on an equal footing was suggested in \cite{KhS}: \beq {\cal L}'_q= -
C_0\left( \frac 1 2 \nu_0 \hat \lambda_{00} \nu_0 + \frac {\gamma} {6 \rho_0} \nu_0^3 +\hat
\lambda_{01} \nu_0 \nu_1 + \hat \lambda_{02} \nu_0 \nu_2 \right), \label{lagrtot}\eeq where
$C_0=(dn/d\eps_{\rm F})^{-1}=\pi^2/(p_{\rm F}m)$  is the usual TFFS normalization factor, the inverse
density of states at the Fermi surface, and $\rho_0=(k_{\rm F}^{0})^3/3\pi^2$   is the density of  one
kind of  nucleon   in equilibrium symmetric nuclear matter. The amplitudes \beq \hat \lambda_{ik} =
\lambda_{ik} + \lambda'_{ik} {\tau}_{1}{\tau}_{2}  \label{isotop} \eeq are the isotopic matrices and
only one of them,\beq \hat \lambda_{00}({\bf r}_1,{\bf r}_2)= \hat \lambda_{00} (1+r_0^2
\triangle_1)\delta({\bf r}_1,{\bf r}_2), \label{lam00} \eeq  is the  finite range operator. The term
proportional to $\gamma$   in equation (\ref{lagrtot})  results in the density dependence of the main,
scalar and isoscalar, Landau--Migdal interaction amplitudes \cite{AB1}.

To minimize the number of new parameters, the ansatz $\lambda'_{01}=\lambda'_{02}=0$ was used in
\cite{KhS}. In this case, the components $\Sigma^{\tau}_1$ and $\Sigma^{\tau}_2$ of the mass operator
do not depend on ${\tau}$, being  functions of the total density  $\nu_0^+=\nu_0^n+\nu_0^p$: \beq
\Sigma_1^{\tau}({\bf r}) = \frac {\delta L_q} {\delta \nu_2^{\tau} ({\bf r})}  = C_0 \lambda_{01}
\nu_0^+ ({\bf r}) \label{Sigm1},\eeq \beq \Sigma_2^{\tau}({\bf r}) = \frac {\delta L_q} {\delta
\nu_2^{\tau} ({\bf r})}  = C_0 \lambda_{02} \nu_0^+ ({\bf r}) \label{Sigm2}.\eeq With the use of
(\ref{rho}) and (\ref{Sigm2}), one can obtain the explicit dependence of the $Z$-factor on the density
with the usual normalization: \beq Z_{\tau}({\bf r})=\frac 2 {1+ \sqrt{1-4C_0 \lambda_{02}\rho^+({\bf
r})/\eps_{\rm F}^0}} \label{Z0}.\eeq

The total interaction energy can be found for the Lagrangian (\ref{lagrtot}) according the canonical
rules. It corresponds to the following EDF: \beq {\cal E}_{\rm int} =   C_0 \left[ \frac 1 2 \hat
\lambda_{00}\left(\nu_0^2 -r_p^2(\nabla \nu_0)^2\right) + \hat \lambda_{01}\nu_0\nu_1 + \frac {\gamma}
{6 \rho_0} \nu_0^3 \right].\label{Energ1} \eeq It does not contain the  `new'  density $\nu_2$ and
converts to the Skyrme EDF at the limit where $\nu_0\to \rho$ and $\nu_1\to \tau$. However, the
replacement of  equation (\ref{rho}) with the $Z$-factor (\ref{Z0}) results in a rather sophisticated
EDF in the self-consistent TFFS, which  can hardly be introduced {\it ad hoc}.

The parameters of the Lagrangian (\ref{lagrtot}) were found in \cite{KhS} by fitting binding energies,
charge radii and single-particle spectra of doubly magic nuclei from $^{40}$Ca to $^{208}$Pb. The
obtained values of $\lambda_{01}=0.31$ and $\lambda_{02}=-0.25$  correspond to the following
characteristics of nuclear matter: $m^*_0=0.95 m$ and $Z_0=0.8$. The latter agrees with the value
found in   \cite{ST1} on the base of the dispersion relation for the quantity $ {\partial \Sigma}/
{\partial \eps}$ \cite{AB1} in nuclear matter.

In \cite{Fay1},  the so-called generalized EDF method was formulated as a generalization of the
Kohn--Sham (KS) method \cite{K-Sh}  for superfluid nuclei. In this case, the EDF depends not only on
the normal densities $ \rho_{\rm n,p}({\bf r})$, but on their anomalous counterparts $ \nu_{\rm
n,p}({\bf r})$ as well. Independently, similar development of the KS method was suggested in condensed
matter physics \cite{Oliveira}. The pairing problem was considered in \cite{Fay1}, with an elegant
method of direct solving Gor'kov equations for spherical systems in the coordinate representation
\cite{Bel}. In practice, this method is close to that of solving HFB equations which was presented
first  in \cite{Gogny} for the Gogny EDF and in \cite{[Dob84]} for the Skyrme EDF.

The KS method is based on the Hohenberg--Kohn theorem \cite{Hoh-K}, stating that the ground state
energy of a Fermi system is a functional of its density. Unfortunately, this theorem does not give any
recipe to construct the EDF. Fayans {\it et al.} \cite{Fay1} found that the EDF (\ref{Energ1}) can be
rather accurately approximated with a rational $\rho$-dependence of equation (\ref{Fay1}) type. In
addition, they used the ansatz $m^*=m$ typical for the KS method. This also agrees  well  with the
above estimation. Thus, the Fayans functional can be interpreted as a simplified version of the
self-consistent TFFS \cite{KhS}, and the  `denominator'  in the EDF (\ref{Fay1}) appears due to the
energy dependence effects taken into account in the TFFS.

 It is worth mentioning that the use of any EDF with density
dependence leads to serious problems if one tries to go beyond mean-field multi-reference
calculations, such as particle number projection or angular momentum projection
\cite{[Dob07d],[Lac09],[Ben09],[Sat11b]}. Therefore, in this work, we use Fayans functional for
single-reference calculations only.

Three sets of the EDF parameters, DF1--DF3, were suggested in  Ref.~\cite{Fay4}, but the most part of
calculations with Fayans EDF were carried out with the set  DF3 \cite{Fay} or its version DF3a for
transuranium region \cite{Tol-Sap}. Although up to now there have been no systematic
 calculations  of nuclear binding energies across the whole nuclear chart  within this
method,  isotopic chains  of spherical nuclei  were  examined in \cite{Fay,Sap-Tol,Tol-Sap}. It was
found that the accuracy is only a little less good than that of the best Skyrme  HFB  calculations. As
for the accuracy of reproducing the charge radii \cite{Sap-Tol} of spherical nuclei, typical deviation
is of the order of 0.01--0.02 fm, i.e. the agreement is on a par with or better than that from Skyrme
EDF models. This may be linked to more adequate density dependence of the Fayans EDF as compared to
the Skyrme one. Indeed, if we denote the average error in describing the binding energies as
$\overline{\delta E}$ and that for the charge radii as $\overline{\delta R_{\rm ch}}$, these
quantities should be, due to the Hohenberg--Kohn theorem \cite{Hoh-K}, proportional to each other:
\beq \overline{\delta R_{\rm ch}}=\alpha \; \overline{\delta E} \label{HK}\,, \eeq where the
coefficient $\alpha$ depends on the functional that we use.
 Often, a fine tuning of the EDF parameters
is performing  by focusing mainly on reproduction of the nuclear masses within a minimal value of
$\overline{\delta E}$. In this case, the accuracy of reproducing the charge radii is proportional to
the coefficient $\alpha$. As the analysis of \cite{Sap-Tol} showed, for the Fayans EDF this
coefficient is less than those of the HFB-17 and SLy4 functionals. Again, this observation may linked
to more enriched density dependence of Fayans functional, which allows to incorporate complex
many-body correlations more efficiently. The Fayans EDF also provides a high quality description of
magnetic \cite{mu1,mu2} and quadrupole \cite{QEPJ,QEPJ-Web}   moments of odd spherical nuclei,
energies and $B(E2)$ values for even semi-magic nuclei \cite{BE2,BE2-Web}. Recent analysis
\cite{Levels} of the single-particle energies of doubly magic nuclei obtained with the Fayans
functional versus the HFB-17 one also provides evidence  in favor of the former.

Up to now, all self-consistent calculations with Fayans functionals were carried out for spherical
nuclei only. In \cite{Tol-Sap}, deformations of the transuranium nuclei were taken into account
approximately. This  work presents the first application of the Fayans functional in studying  axially
deformed nuclei.

\section{FaNDF$^0$ functional}
For completeness, we write down explicitly  main ingredients of the Fayans EDF method. In this method,
the ground state energy of a nucleus is considered as a functional of normal and anomalous densities,
\beq E_0=\int {\cal E}[\rho({\bf r}),\nu({\bf r})] d^3r,\label{E0} \eeq  where the isotopic indices
and the spin-orbit densities are for brevity omitted.

The main distinctions between this method and the Skyrme EDF approach  lie  in the normal part of the
EDF ${\cal E}_{\rm norm}$,  containing  the central  and spin-orbit  terms,  and the Coulomb
interaction term for protons. In most applications of this method \cite{Fay3,Fay4,Fay}, the DF3
functional was used with the finite range Yukawa-type  central force. In this  work  we use the EDF
FaNDF$^0$ from \cite{Fay5} with a localized form of the Yukawa function, ${\rm Yu}(r)\to 1-r_c^2
\nabla^2$, which makes the structure of the surface part of the EDF closer to that from the Skyrme
functionals. This form allows us to use a modified version of the computer code \textsc{HFBTHO}
\cite{code}, originally constructed for Skyrme-like EDFs. The parameters of FaNDF$^0$ were fitted to
the EOS of nuclear and neutron matter by Friedman and Pandharipande \cite{Pandh} and the masses of
lead and tin isotopes.

The volume part of the EDF, ${\cal E}^{\rm v}(\rho)$, is taken  as a fractional function of densities
$\rho_+=\rho_n+\rho_p$ and $\rho_-=\rho_n-\rho_p$: \beq {\cal E}^{\rm v}(\rho)=C_0 \left[ a^{\rm
v}_+\frac{\rho_+^2}4 f^{\rm v}_+(x) + a^{\rm v}_-\frac{\rho_-^2}4 f^{\rm v}_-(x)\right], \label{EDF_v}
\eeq where \beq f^{\rm v}_+(x)=\frac{1-h^{\rm v}_{1+}x^{\sigma}}{1+h^{\rm v}_{2+}x^{\sigma}}
\label{fx_vpl} \eeq and \beq f^{\rm v}_-(x)=\frac{1-h^{\rm v}_{1-}x}{1+h^{\rm v}_{2-}x}.
\label{fx_vmi} \eeq Here, $x=\rho_+/\rho_0$ is the dimensionless nuclear density. The power parameter
$\sigma=1/3$ is chosen in the FaNDF$^0$ functional, in contrast to the case for DF3, where $\sigma=1$
is used. The structure of other terms in the volume parts of these two functionals is kept the same.
However, the above   difference leads to significantly different values of the dimensionless
parameters in equations (\ref{EDF_v})--(\ref{fx_vmi}) although they still correspond to the same
characteristics of nuclear matter, the  incompressibility  $K_0=220$ MeV, equilibrium density
$\rho_0=0.160$ fm$^{-3}$ ($r_0=1.143$ fm), and  energy per particle $\mu=-16.0$ MeV.  The parameters
denoted by `+' are  $a^{\rm v}_+=-9.559$, $h^{\rm v}_{1+}=0.633$, $h^{\rm v}_{2+}=0.131$,  and the
parameters denoted by `-'  are $a^{\rm v}_-=4.428$, $h^{\rm v}_{1-}=0.25$, $h^{\rm v}_{2-}=1.300$,
which all are dimensionless quantities.  This parameter set corresponds to the asymmetry energy
coefficient  of  $a_{\rm sym}=30.0$ MeV.

 In \cite{criteria}, a set of criteria were suggested for the Skyrme EDFs.
For the nuclear matter part these criteria were connected to properties of the saturation point and to
the second derivatives of the energy density ${\cal E}^{\rm v}(\rho_+,\rho_-)$. Using the notations of
\cite{criteria}, the so-called skewness coefficient of the symmetrical nuclear matter is equal to \beq
Q_0 = 27  \rho_0^3 \left( \frac {\partial^3 {\cal E}(\rho_+, \rho_-)/\rho_+ } {\partial \rho_+^3}
\right)_{x=1,y=0}\, , \ \label{Q0} \eeq  where  $y=\rho_-/\rho_0$ is a dimensionless neutron excess.
To simplify the expressions for the mixed higher density derivatives, we introduce a nuclear matter
energy function \beq S(\rho_+) = \frac 1 2  \rho_+^2 \left( \frac {\partial^2 {\cal E}(\rho_+,
\rho_-)/\rho_+ } {\partial \rho_-^2} \right)_{y=0}\,. \ \label{Snm} \eeq The first and second density
derivatives of this function are \beq L_0  = 3 \rho_0   \left( \frac {\partial {\cal S}(\rho_+) }
{\partial \rho_+} \right)_{x=1}\, , \ \label{L0} \eeq and  \beq K_{\rm sym} = 9 \rho_0^2 \left( \frac
{\partial^2 {\cal S}(\rho_+) } {\partial \rho_+}^2 \right)_{x=1}\, . \ \label{Ksym} \eeq The numerical
values  of the above quantities for the FaNDF$^0$ functional are presented in table \ref{tab0} where
they are compared to the Skyrme EDFs taken from \cite{criteria}, which are used in the present work.
One can see that the differences between all of the listed main nuclear matter characteristics are
usually small.

\begin{table}
\caption{\label{tab0} Nuclear matter characteristics for differen EDFs, all of which are given in
 MeV except the matter density $\rho_0$.}
\begin{indented}
\bigskip

\item[]\begin{tabular}{|c|c|c|c|c|c|c|c|} \hline
EDF      & $\rho_0$, fm$^{-3}$ & $\mu$ & $a_{\rm sym}$ & $K_0$ & $Q_0$ & $L_0$ & $K_{\rm sym}$  \\
\hline
FaNDF$^0$&0.160     & -16.00 & 30.00  & 220.00  & -427.14  &  29.96   & - 149.22    \\
SkM*     &0.160     & -15.77 & 30.03  & 216.61  & -386.09  &  45.78   & - 155.94    \\
SLy4     &0.160     & -15.97 & 32.00  & 229.91  & -363.11 &  45.94   & -119.73   \\
SLy6     &0.159     & -15.92 & 31.96  & 229.86  & -360.24 &  47.45   & -112.71  \\

\hline
\end{tabular}
\end{indented}
\end{table}

The main difference between FaNDF$^0$ and DF3 functionals lies in the structure of the surface term.
Now it is as follows: \beq {\cal E}^{\rm s}(\rho)=C_0 \frac 1 4 \frac { a^{\rm s}_+ r_0^2(\nabla
\rho_+)^2}{1+h^{\rm s}_{+}x^\sigma+h^{\rm s}_\nabla r_0^2(\nabla x_+)^2}, \label{EDF_s}\eeq with
$h^{\rm s}_{+}=h^{\rm v}_{2+}$, $a^{\rm s}_+=0.600$, $h^{\rm s}_\nabla=0.440$.

 The usual form for the direct Coulomb term of the EDF  of \cite{Fay5} is employed;
the  folded  charge density $\rho_{\rm ch}$ is found  taking into account the proton and neutron form
factors. As regards the exchange Coulomb term, it was taken as follows: \beq -\frac 3 4\left(\frac 3
\pi\right)^{1/3}e^2\rho^{4/3}_{\rm p}(1-h_{\rm Coul}x^\sigma_+), \eeq with $h_{\rm Coul}=0.941$. Such
a strong suppression, in comparison with the Slater approximation with $h_{\rm Coul}=0$, helps with
solving the so-called Nollen--Schiffer anomaly \cite{N-Sch}. It is worth mentioning that a similar
suppression of the Coulomb exchange term was adopted in some
 Skyrme functionals \cite{SkX}.

The usual  form  for the TFFS \mkrm{structure of the} spin-orbit term  \cite{AB1} was used in
FaNDF$^0$ with the same spin-orbit parameters as in the DF3 functional \cite{Fay}.  Note that the
FaNDF$^0$ functional does not contain the effective tensor, in contrast to the DF3 \cite{Fay3} and
DF3a \cite{Tol-Sap} EDFs.

 For completeness, we write out explicitly the anomalous
term of the EDF \cite{Fay5}:  \beq {\cal E}_{\rm anom}= C_0 \sum_{i=n,p} \nu_i^{\dag}({\bf r})
f^{\xi}(x_+({\bf r}))\nu_i({\bf r}), \label{Eanom}\eeq where the density-dependent dimensionless
effective pairing force is \beq f^{\xi}(x_+)= f^{\xi}_{\rm ex} + h^{\xi} x_+ + f^{\xi}_{\nabla}
r_0^2(\nabla x_+)^2, \label{fksi}\eeq with $f^{\xi}_{\rm ex}=-2.8$, $h^{\xi}=2.8$,
$f^{\xi}_{\nabla}=2.2$.

All the above values of the parameters were found in  Ref.~\cite{Fay5} by fitting the masses and
charge radii of approximately a hundred spherical nuclei, from calcium  isotopes  to lead  isotopes.
 In this work, we use this same parameter set for deformed nuclei.

In our current implementation of the Fayans functional in the computer code \textsc{HFBTHO}, for
technical reasons  we made two small simplifications
 to  the original   FaNDF$^0$  EDF . Firstly, we used the
approximation $\rho_{\rm ch}=\rho_{\rm p}$ for the direct Coulomb term. Secondly, we put
$f^{\xi}_{\nabla}=0$ in (\ref{fksi}) making the anomalous EDF closer to that used in \cite{Fay}.
Therefore,  below we use the above parameters for the normal part of the EDF only. As regards the
anomalous EDF, the parameters will be found anew and will be given in the corresponding places. As
long as we are dealing with the zero-range pairing force, the strength parameters depend on the cutoff
energy $E_{\rm cut}$ in the pairing problem, being also smoothly $A$ dependent \cite{Fay}. In
practice, this means that we take $f^{\xi}=-0.440$ for U and $f^{\xi}=-0.448$ for Pb isotopes.

\section{Results}
\subsection{The uranium chain}
We  have chosen  uranium isotopes for the first application of the FaNDF$^0$  EDF to deformed nuclei,
 since  most of them have a well established stable deformation.  There have been numerous
Skyrme EDF studies concerning the deformation landscape of actinide nuclei; see e.g.
\cite{Gor09,Sta09,Kort1,McDon1}, to list but a few recent studies.

In this  work, due to  the axial computer code employed, we limit ourselves to the quadrupole
deformation $\beta_2$ only, with reflection symmetry assumed. We  have mainly  focused on the ground
state characteristics which are,  unlike  the fission barriers, within the reach of the current axial
framework. Indeed, the ground states of U isotopes are expected to be axially deformed. Triaxial
deformation, which  usually appears at the top of the fission barriers,  is neglected  in  the present
study.  Also, the role of octupole degrees of freedom becomes important around the second fission
barrier in the case of asymmetric fission. For the ground states of uranium isotopes,  we present a
systematic comparison of our results with  those obtained with two versions of Skyrme HFB functionals:
HFB-17 and HFB-27 \cite{site}.

\begin{figure}
\centerline {\includegraphics [width=80mm]{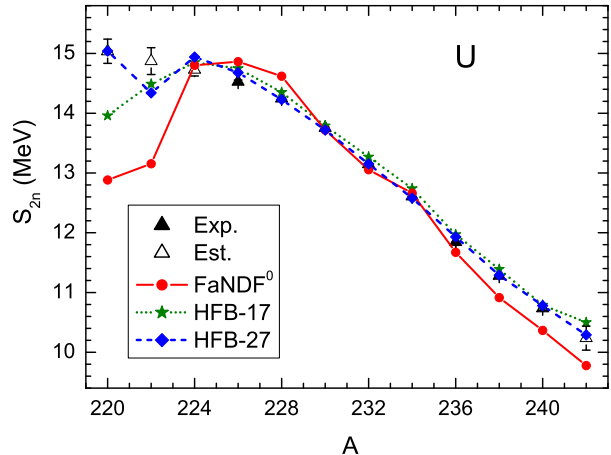}} \vspace{2mm} \caption{Two-neutron separation
energies $S_{2n}$ for even U isotopes. Predictions from the FaNDF$^0$ functional are compared with
those from two Skyrme EDFs: HFB-17 and HFB-27. Empty triangles show the estimated values
\cite{mass}.}\label{fig:US2n}
\end{figure}

For the uranium chain, we found that our results converged when the number of oscillator shells was
equal to $N_{\rm sh}=25$, i.e. the change of this number to $N_{\rm sh}=30$ practically does not
influence the results. As regards the pairing force, the set of \cite{Fay5} with $f^{\xi}_{\rm
ex}=-h^{\xi}$ corresponds to the `surface' pairing with a strong attraction at the surface and very
small value of $f^{\xi}$ inside a nucleus. Such a model of pairing is typical for all versions of the
Fayans functional \cite{Fay,BE2}. Here we  have  found that the deformation energy for surface pairing
is very close to that for the `volume' pairing model (corresponding to  $h^{\xi}=0$), provided the
deformation parameter $\beta_2$ is less than  0.3--0.5. To stress the effect of the specific density
dependence of the normal part of the Fayans EDF, we use  as a rule  the simplest one-parameter volume
pairing. The cutoff energy $E_{\rm cut}=60\;$MeV is chosen, with the corresponding value
$f^{\xi}=-0.440$ fitted to the double mass differences for the uranium isotopes.

In figure \ref{fig:US2n}, two-neutron separation energies \beq S_{2n}(N,Z)=B(N,Z)-B(N-2,Z),
\label{S2n} \eeq  are shown for uranium isotopes.  Comparison is made with experimental data
\cite{mass} and predictions from the  HFB-17 and HFB-27 EDFs. Taking into account that the parameters
of the FaNDF$^0$ functional were fitted only for spherical nuclei  not heavier than that of lead, the
description of $S_{2n}$ values for uranium isotopes looks rather reasonable. The deviation  of
0.5\,MeV from the experimental $S_{2n}$ values for heavy U isotopes is explained mainly by two
features. The first feature is the use of a simple volume pairing interaction. The second feature is
absence of the effective tensor term in the FaNDF$^0$ EDF. Indeed, as was shown in \cite{Tol-Sap}, the
tensor term is especially important in uranium and transuranium region as, in the spherical case, high
$j$ levels dominate in vicinity of the Fermi level for these nuclei. As a result, the spin-orbit
density, which comes to the EDF together with the tensor force, is typically large in these nuclei,
changing significantly along the isotopic chain. Accounting for these effects, with the tensor force,
represents the essential difference between the DF3a and DF3 EDFs.

\begin{figure}
\centerline {\includegraphics [width=80mm]{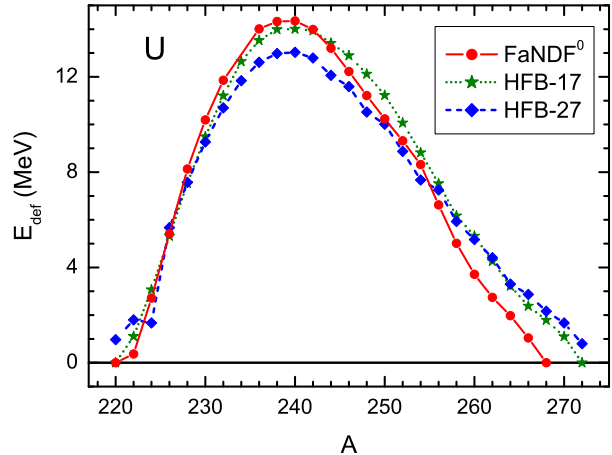}} \vspace{2mm} \caption{Deformation energy $E_{\rm
def}$ for even U isotopes. Predictions from the FaNDF$^0$ functional are compared with those from two
Skyrme EDFs: HFB-17 and HFB-27.}\label{fig:UEdef}
\end{figure}

\begin{figure}
\centerline {\includegraphics [width=80mm]{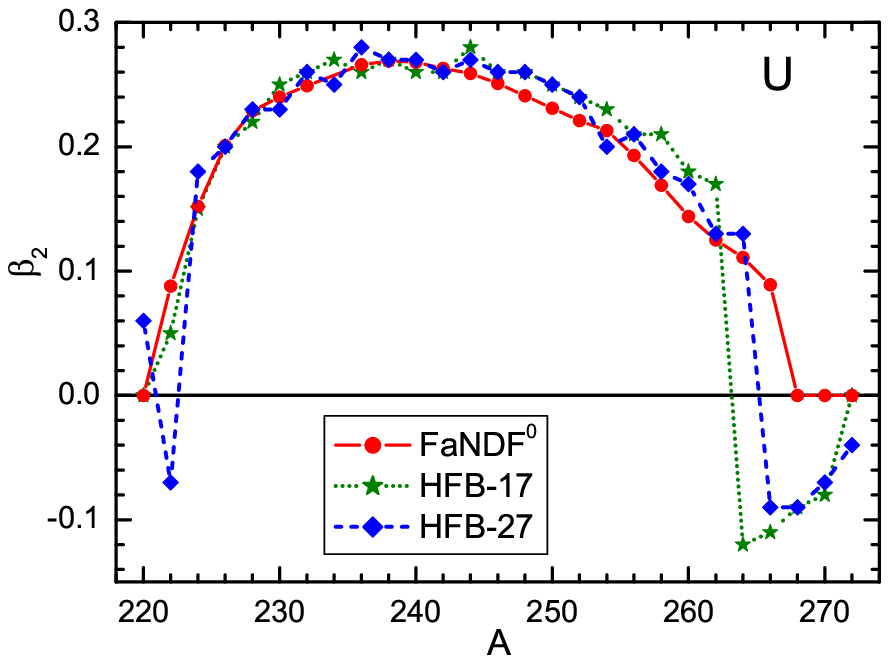}} \vspace{2mm} \caption{Quadrupole deformation
parameter for even U isotopes. Predictions from the FaNDF$^0$ functional are compared with those from
two Skyrme EDFs: HFB-17 and HFB-27.}\label{fig:Ubet}
\end{figure}

\begin{figure}
\centerline {\includegraphics [width=80mm]{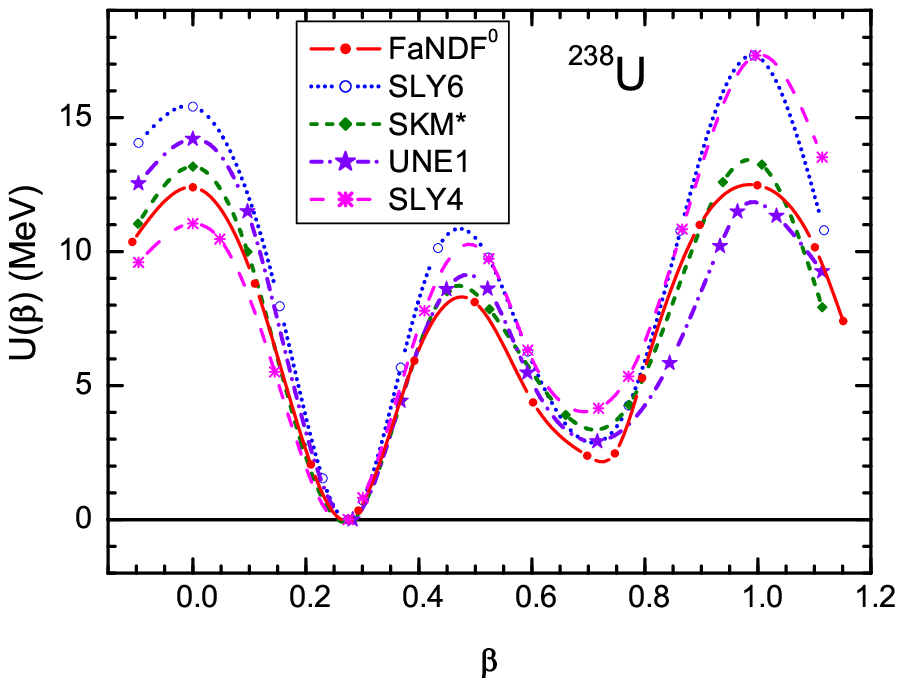}} \vspace{2mm} \caption{The deformation energy curve
for  $^{238}$U. Predictions from the FaNDF$^0$ functional are compared with those from four Skyrme
EDFs. UNE1 label refers to the UNEDF1 EDF.}\label{fig:U238}
\end{figure}

In figures \ref{fig:UEdef} and \ref{fig:Ubet}, a comparison is presented to the same Skyrme
functionals for the deformation energy: \beq E_{\rm def}(\beta_2) = B(\beta_2) - B(\beta_2=0),
\label{Edef} \eeq and the deformation parameter itself. Unfortunately, both  of  these  quantities
have no direct experimental equivalent. We see that our calculations with the FaNDF$^0$ functional
agree reasonably with both of the Skyrme EDF predictions.

To examine  the applicability  of the FaNDF$^0$ functional for the description of large deformations,
we  have  calculated the deformation energy curve for the $^{238}$U nucleus  up to the  second
 fission  barrier,  shown in  figure \ref{fig:U238}.
 The calculation scheme remains the same as in the above consideration of the ground states, with
a one exception: we have replaced the more simpilied volume pairing interaction with a surface
pairing, with parameter values $f^{\xi}=-1.433,\, h^{\xi}=1.375$. This kind of pairing force has been
found to be more realistic \cite{Fay5,Fay,BE2}; see also the {\it ab initio} consideration of the
pairing force in \cite{Pankrat1}. The difference between the surface and volume pairings becomes
important for high deformations, as the role of the surface is strengthened in this case. For example,
the second barrier in figure \ref{fig:U238} will be 2 MeV higher if we take the volume pairing. For
comparison, we  also show the results obtained with SLy6 \cite{sly4to7}, SkM* \cite{skms}, UNEDF1
\cite{Kort1}, and SLy4 \cite{sly4to7} Skyrme EDFs.  All calculations  were  carried out within the
same calculation scheme  as for the FaNDF$^0$ functional, i.e. with account taken of just the
quadrupole deformation, without triaxiality or octupole degrees of freedom. On neglecting these
degrees of freedom, the calculated inner fission barries are typically raised by a few MeV and the
outer fission barries, in asymmetric fission, by substantially more; see e.g. \cite{Bur04}. Therefore,
it is not meaningful to compare numerical values of the barriers in figure \ref{fig:U238} with
experimental values directly. However, it is worth noting that the FaNDF$^0$ curve is rather close to
the SkM* and UNEDF1 ones which both, particularly the UNEDF1 one, after the inclusion of triaxial and
octupole degrees of freedom, describe uranium barriers reasonably well \cite{Kort1}. On the  basis  of
the results obtained for the uranium isotopes, it seems reasonable to apply  this functional for the
analysis of the deformation characteristics of other isotopic chains.

\begin{figure}
\centerline {\includegraphics [width=80mm]{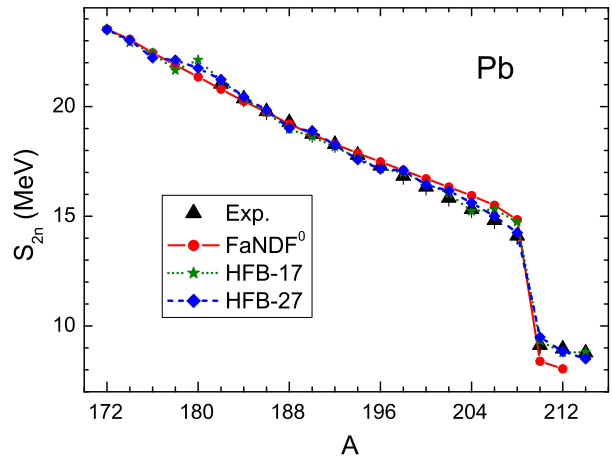}} \vspace{2mm} \caption{Two-neutron separation
energies $S_{2n}$ for even Pb isotopes. Predictions from the FaNDF$^0$ functional are compared with
those from two Skyrme EDFs: HFB-17 and HFB-27.}\label{fig:PbS2n}
\end{figure}

\begin{figure}
\centerline {\includegraphics [width=80mm]{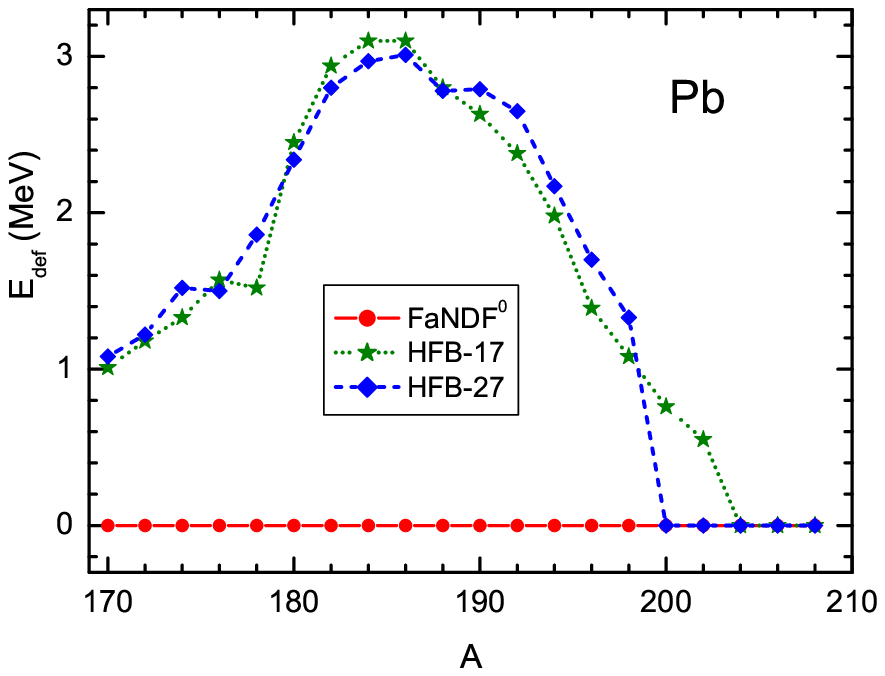}} \vspace{2mm} \caption{Deformation energy $E_{\rm
def}$ for even Pb isotopes. Predictions from the FaNDF$^0$ functional are compared with those from two
Skyrme EDFs: HFB-17 and HFB-27.}\label{fig:PbEdef}
\end{figure}

\begin{figure}
\centerline {\includegraphics [width=80mm]{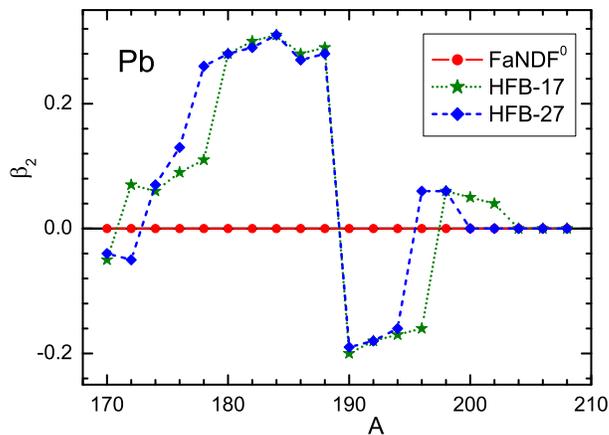}} \vspace{2mm} \caption{Quadrupole deformation
parameter for even Pb isotopes. Predictions from the FaNDF$^0$ functional are compared with those from
two Skyrme EDFs: HFB-17 and HFB-27.}\label{fig:Pbbet}
\end{figure}

\subsection{The lead chain}
Our interest to the lead chain is motivated by the observation that HFB-27 and other functionals of
this family predict \cite{site} rather strong deformations, $|\beta_2|\simeq$ 0.2--0.3, for light
$^{180-192}$Pb isotopes.  Also, many other Skyrme EDFs seem to predict, at the mean-field level, the
appearance of deformation in the region of light Pb isotopes; see e.g. \cite{Sto03,Erl12}.  In our
opinion,  this does not correspond to the trend of the  empirical data on charge radii \cite{Sap-Tol}
 or  magnetic moments \cite{mu1,mu2}. Indeed, charge radii produced by the HFB-27
EDF \cite{site} describe the data for heavy Pb isotopes perfectly well but disagree significantly for
those lighter than $^{192}$Pb.  Analysis of \cite{Robledo} within the generator coordinate method with
the use of the Gogny force D1S and of \cite{Bender} with the SLy6 EDF confirmed the spherical form for
these neutron-deficient Pb isotopes. Although in both studies the angular momentum projection
technique was used, the simplest mean-field calculation also did not predict so large deformations as
in \cite{site}. Thus predictions of different Skyrme functionals for the light Pb isotopes are
essentially different. Therefore it is of interest to look at how the FaNDF$^0$ functional behaves for
these nuclei and compare our predictions with those of various Skyrme EDFs.

In the analysis of the lead isotopes, we use the same calculation scheme as for the uranium chain,
i.e. $N_{\rm sh}=25$ and $E_{\rm cut}=60\;$MeV are chosen, and  the corresponding value
$f^{\xi}=-0.448$ is a bit different in order to provide a better description of the  $S_n$ and
$S_{2n}$ values on average. We again begin with the two-neutron separation energies $S_{2n}$, shown in
figure \ref{fig:PbS2n}. On average, the agreement with the data obtained with FaNDF$^0$  is only a
little less good compared to the case for the HFB-17 and HFB-27 functionals. The agreement of the
$S_{2n}$ values with the experiment for the heaviest Pb isotopes will be better if we use the surface
pairing, which is more realistic as is discussed above. However, a new readjustment of FaNDF$^0$
parameters with tensor terms added is necessary for competing with HFB EDFs in the accuracy of
reproducing the binding energies.

As regards the deformation characteristics, there is a  notable  difference between predictions from
the Fayans  FaNDF$^0$ EDF and those from the two Skyrme functionals under consideration. Namely,
calculations with the Fayans functional result  to a  spherical  mean-field   ground state  for  all
of the lead isotopes. At the same time, the HFB-17 and HFB-27 functionals both predict a stable
deformation in the ground states  for  many light Pb isotopes, as shown in figure \ref{fig:Pbbet}. For
the HFB-27 functional, deformation appears for isotopes with $A=$ 170--198,  and  for the HFB-17
functional, for all isotopes with $A<204$. For both the functionals, the deformation changes sign from
positive for $^{188}$Pb to negative for $^{190}$Pb, and the deformation is strong for isotopes with
$A=$ 180--192; $\beta_2 \simeq 0.3$ for $A=$ 180--188  and $\beta_2 \simeq - 0.2$ for $A=$ 190--194 in
the case of the HFB-27 functional and for $A=$ 190--196 in the case of the HFB-17 one. Thus,
 the value of the  deformation parameter  within this  mass-region  is  approximately of the
same order of magnitude as for  the uranium isotopes. The deformation energy is less, $E_{\rm
def}\simeq 3\;$MeV, and also not  negligible,  as shown in figure \ref{fig:PbEdef}.  To summarize, the
predictions of both of these  HFB  functionals  do not follow   experimental data  trends  on charge
radii and magnetic moments, as was discussed above.  In addition,  they  disagree with  the
predictions of \cite{Bender} for the SLy6 EDF.

\begin{figure}
\centerline {\includegraphics[width=0.9\linewidth]{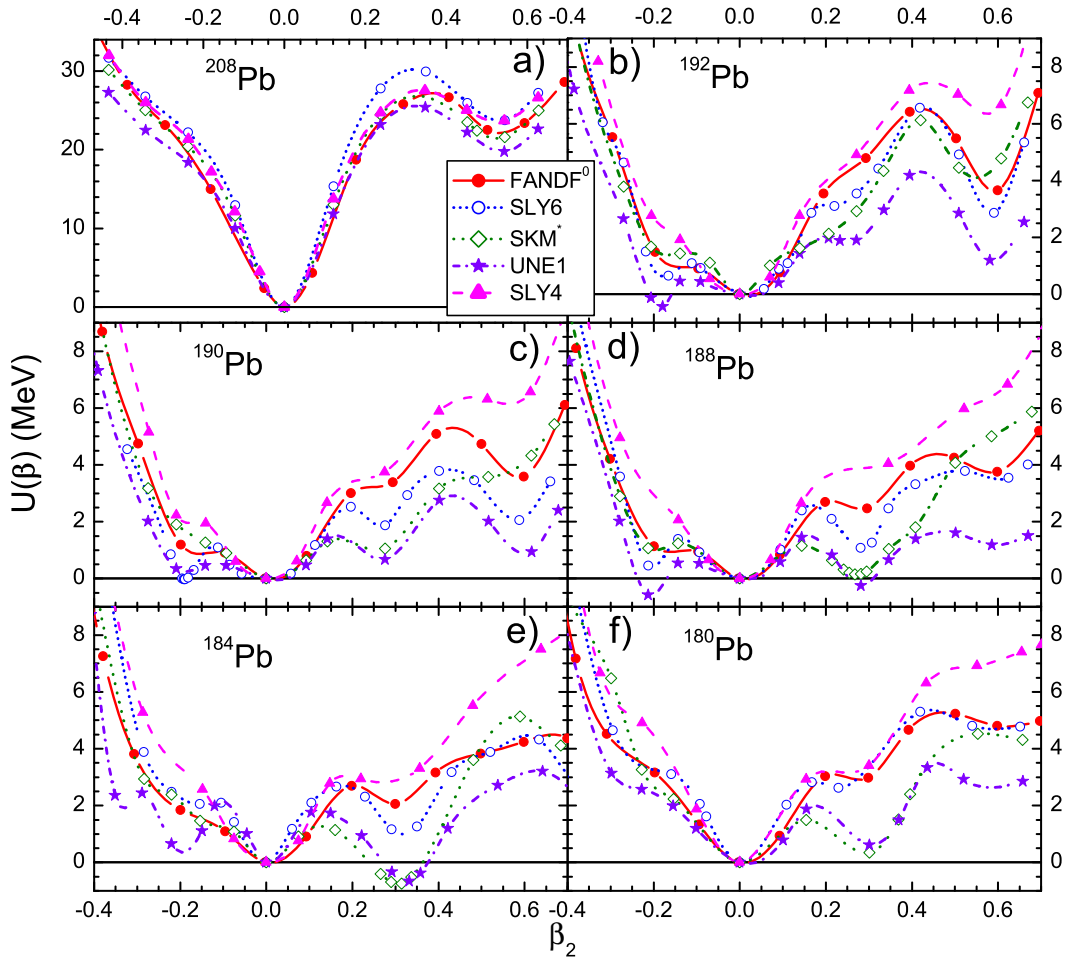}} \caption{Deformation energy curves
$U(\beta)$ for Pb isotopes as a function of the deformation $\beta$ for FaNDF$^0$ and various Skyrme
EDFs. Shown are the results for $^{208}$Pb (a), $^{192}$Pb (b), $^{190}$Pb (c), $^{188}$Pb (d),
$^{184}$Pb (e), and $^{180}$Pb (f).}\label{fig:Pbcurves}
\end{figure}

To investigate the problem in more detail, we  have  calculated the deformation energy curves
$U(\beta)$ for several light Pb isotopes with the FaNDF$^0$ functional and the same Skyrme EDFs  as
 for the case of $^{238}$U. The results are shown in figure~\ref{fig:Pbcurves}.  We begin
 comparison  with the doubly  magic $^{208}$Pb, shown in panel (a). All
four curves behave in a similar way, which corresponds to the very rigid nature of this nucleus. The
positions of the first barrier and the second minimum are almost the same for all of  the EDFs. For
the small deformation region,  FaNDF$^0$ and UNEDF1 curves  show very similar behavior.  The SkM*
 deformation energy curve seems to be the one closest  to FaNDF$^0$,  and only SLy6 deformation
energy is  notably  higher, by  4--5 MeV at $\beta_2=$ 0.3--0.6.

 Next, we investigate the light isotopes, which are of our main interest  in the present work.
For $^{192}$Pb nucleus,  shown in  figure \ref{fig:Pbcurves},  panel (b), the FaNDF$^0$
 deformation energy follows the SLy6 and SkM* ones rather closely,  being rather
rigid  at the minimum of $\beta=0$. Only the UNEDF1 curve behaves softer predicting an oblate,
$\beta_2\simeq -0.2$, ground state. The SLy4 EDF is the most rigid. The latter is true also for the
$^{190}$Pb nucleus,  shown in panel (c).  The  behavior of the FaNDF$^0$ curve  here  is also rather
rigid, but softer  compared to $^{192}$Pb. It has a shallow minimum,  roughly  $1\;$MeV  above the
ground state energy, at $\beta_2\simeq$ -(0.1--0.2). According to the remaining  three Skyrme EDF
predictions, this nucleus is much softer. All of the corresponding Skyrme functions possess clearly
distinguishable minima at the prolate deformation, $\beta_2\simeq 0.3$. The corresponding
 excitation energies  are about 2 MeV for SLy6, 1 MeV for SkM* and only 0.5 MeV for the UNEDF1
EDF. For oblate deformations,  UNEDF1 and SLy6 EDFs lead to very low minima at $\beta_2\simeq -0.2$,
the latter being a little lower than the spherical one.  As mentioned, these predictions are based on
the single-reference mean-field level.  Corrections to this scheme leads to a  restoration  of the
spherical form for this nucleus \cite{Bender}.

\begin{figure}
\centerline {\includegraphics[width=0.5\linewidth]{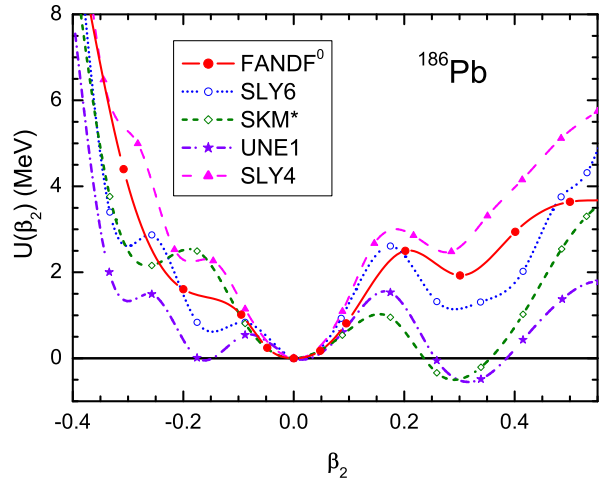}} \caption{Deformation energy curves
$U(\beta)$ for  the  $^{186}$Pb nucleus. }\label{fig:Pb186}
\end{figure}

For the $^{188}$Pb nucleus,  shown in panel (d) of  figure \ref{fig:Pbcurves}, the situation is
similar, but now both  of  the deformed UNEDF1 minima are lower  compared to  the spherical one. The
SkM* minimum is higher than the spherical one, but only a bit. The FaNDF$^0$ curve is qualitatively
similar to the SLy6 one, but a little more rigid. For the $^{184}$Pb,  shown in panel (e), SkM* and
UNEDF1 EDFs predict  a prolate ground state at $\beta_2\simeq 0.3$. Finally, the $^{180}$Pb nucleus,
 shown in panel (f), becomes much more rigid than the nuclei considered above.

 In figure \ref{fig:Pb186}, we show separately the deformation energy curves for the $^{186}$Pb
nucleus possessing, in addition to the ground state, two exited low-lying 0$^+$ states, an oblate and
a prolate one, with excitation energy of about 1 MeV. As shown in \cite{Robledo} and \cite{Bender}, it
is necessary to go beyond the plain mean field theory to describe their characteristics correctly.
Here, the mean field picture gives a rough estimate which of the EDFs has the better chance  of
providing a successful description of these states in  calculations beyond the mean field level. In
\cite{Bender}, SLy6 was found to reproduce these state successfully. Here, the FaNDF$^0$ curve is
again qualitatively similar to SLy6, but both minima are approximately 1 MeV higher.

To conclude this section, we note that predictions from different Skyrme EDFs  for light lead isotopes
are  found to be  quite different. The SLy4 EDF is the most rigid of all functionals under
consideration, including FaNDF$^0$.  On the other hand,  FaNDF$^0$ predicts spherical form for all the
lead isotopes.   Probably, this could  be explained with the influence of the denominator of equation
(\ref{EDF_v}), which provides some feedback preventing the deformation of the light Pb isotopes. At
the same time, FaNDF$^0$ predictions are rather close to SLy6 ones, with exception of those for the
$^{190}$Pb nucleus.

\section{Conclusion}
This  work   presents the first application of the Fayans functional FaNDF$^0$ \cite{Fay5} to deformed
nuclei. The Fayans functional makes an interesting alternative to the Skyrme EDF with some promising
properties, as shown in the current work.  A systematic comparison for  the mean-field  deformations
and deformation energies was made against two modern Skyrme EDFs: HFB-17 and HFB-27. Results were
calculated for the uranium  and lead isotopic chains. In the uranium case, our results are
qualitatively close to both  the HFB-17 and HFB-27 functional results.

To check the applicability  of the Fayans functional for description of large deformation, we
calculated the deformation energy curve for $^{238}$U nucleus  with FaNDF$^0$  and four different
Skyrme EDFs. Our result turned out to be rather close to the SkM* and UNEDF1 ones. These two Skyrme
parameterizations, in particularly UNEDF1,  after inclusion of the octupole deformation  and
triaxiality, reproduce values of the experimental first and second barriers in this nucleus
sufficiently well \cite{Kort1}. Here, in the present work, we limit ourselves to just the quadrupole
deformation only  due to limitations of the used computer code. For the  $^{238}$U,  the results
obtained in axial framework for  FaNDF$^0$ are rather close to those of most successful Skyrme EDFs.
Nevertheless, with the current calculation scheme, it is too early to draw any concrete conclusions
about the fission properties of FaNDF$^0$.

 For the lead isotopes, however, there was some notable differences between FaNDF$^0$ and Skyrme EDFs.
Here, both of the HFB functionals predict strong deformation of the light isotopes: $A=$ 178--196 for
the functional HFB-17 and $A=$ 178--194 for HFB-27. This  does not agree to experimental data on the
charge radii \cite{Sap-Tol} and magnetic moments \cite{mu1,mu2}. On the contrary, the Fayans
functional predicts spherical  mean-field solution  for all Pb isotopes, in agreement with
experimental trend.  To examine these  differences, we calculated deformation energy curves
 for several light lead isotopes. Again the FaNDF$^0$ results are compared to those
obtained with four Skyrme EDFs.  The predictions from the different Skyrme EDFs are quite different,
the FaNDF$^0$ ones being rather close to those from the SLy6 EDF.

Thus, the FaNDF$^0$ functional, with the parameters adjusted  to spherical nuclei,  seems to  describe
rather well the  ground state  deformation properties of the two isotopic chains studied in the
present work. This feature may be linked to a peculiar density dependence of the Fayans functional,
resulting from the energy dependence effects of the self-consistent TFFS \cite{KhS} which are hidden
in the formulation in terms of the EDF. A systematic analysis of deformed nuclei with the Fayans
functional would be necessary to estimate its possible benefits across larger portions  of the nuclear
chart. Also, to address the fission properties of the FaNDF$^0$, fully triaxial calculations are
required.

\section{Acknowledgments}
 We are grateful to Michael Bender for useful comments.
The work was partly supported  by the Grant NSh-932.2014.2 of the Russian Ministry for Science and
Education, by the RFBR Grants  12-02-00955-a, 13-02-00085-a, 13-02-12106\_ofi-m, 14-02-00107-a,
14-22-03040\_ofi-m, the Grant by IN2P3-RFBR under Agreement No. 110291054, and Swiss National
Scientific Foundation Grant No. IZ73Z0\_152485 SCOPES.   This work was also supported (M.K.) by the
Academy of Finland under the Centre of Excellence Programme 2012--2017 (Nuclear and Accelerator Based
Physics Programme at JYFL) and the FIDIPRO programme; and by the European Unions Seventh Framework
Programme ENSAR (THEXO) under Grant No. 262010.

{}

\end{document}